\pdfoutput=1

\documentclass{amsart}

\usepackage{amsmath}    
\usepackage{amssymb}    
\usepackage{dsfont} 	
\usepackage{bm}         
\usepackage{mathtools}  
\usepackage[hypertexnames=false]{hyperref} 
\usepackage{enumerate}
\usepackage{multirow}
\usepackage{array}
\usepackage{rotating}
\usepackage{lscape}
\usepackage{pdflscape}
\usepackage{longtable}
\usepackage{cite}
\usepackage{color}
\usepackage[english]{babel}
\usepackage{setspace}

\makeatletter
\newcommand{\addresseshere}{%
  \enddoc@text\let\enddoc@text\relax
}

\newcommand{\vet}[1]{{\ensuremath{\mbox{\boldmath $#1$}}}}
\def\ci{\perp\!\!\!\perp}

\begin{document}

\title[Pathophysiological Domains Underlying the Metabolic Syndrome]
{Pathophysiological Domains Underlying the Metabolic Syndrome: An Alternative
Factor Analytic Strategy}

\author[C.F.W.\ Peeters]{Carel F.W.\ Peeters*}
\thanks{*Corresponding author.}
\address[Carel F.W.\ Peeters]{
Dept.\ of Epidemiology \& Biostatistics \\
VU University medical center Amsterdam \\
Amsterdam\\
The Netherlands}
\email{cf.peeters@vumc.nl}

\author[J.\ Dziura]{James Dziura}
\address[James Dziura]{
Yale Center for Analytical Sciences \\
Yale School of Public Health \\
New Haven, CN, USA}
\email{james.dziura@yale.edu}

\author[F.\ van Wesel]{Floryt van Wesel}
\address[Floryt van Wesel]{
Dept.\ of Methodology \& Statistics \\
Utrecht University \\
Utrecht\\
The Netherlands}
\email{F.vanWesel@$\mathrm{uu}$.nl}

\begin{abstract}\label{abstract}
Factor analysis (FA) has become part and parcel in
metabolic syndrome (MBS) research. Both exploration- and
confirmation-driven factor analyzes are rampant. However, factor
analytic results on MBS differ widely. A situation that is at least
in part attributable to misapplication of FA. Here, our purpose is (i) to review
factor analytic efforts in the study of MBS with emphasis on misusage
of the FA model and (ii) to propose an alternative factor analytic strategy.

\noindent\emph{Methods:} The proposed factor analytic strategy
consists of four steps and confronts weaknesses in
application of the FA model. At its heart lies the explicit
separation of dimensionality and pattern selection as well as the direct
evaluation of competing inequality-constrained loading patterns.
A high-profile MBS data set with anthropometric measurements on
overweight children and adolescents is reanalyzed using
this strategy.

\noindent\emph{Results:} The reanalysis
implied a more parsimonious constellation of pathophysiological
domains underlying phenotypic expressions of MBS than the original
analysis (and many other analyzes). The results
emphasize correlated factors of impaired glucose metabolism and
impaired lipid metabolism.

\noindent\emph{Conclusions:}
Pathophysiological domains underlying phenotypic expressions of MBS included in the
analysis are driven by multiple interrelated metabolic impairments. These
findings indirectly point to the
possible existence of a multifactorial aetiology.

\begin{sloppypar}
\bigskip \noindent \footnotesize {\it Key words}: Factor analysis;
Metabolic syndrome

\bigskip
\noindent \emph{Abbreviations}: AACE = American Academy of Clinical
Endocrinologists; BMI = body mass index; CFA = confirmatory factor
analysis; CVD = cardiovascular disease; DBP = diastolic blood
pressure; DM = diabetes mellitus; EFA = exploratory factor analysis;
EGIR = European Group for study of Insulin Resistance; FA = factor
analysis; G2 blood glucose level two hours after oral glucose
intake; GB = blood glucose level at (fasting) baseline; HDL chol. =
high density lipoprotein cholesterol; HOMA = homeostatic model
assessment; IR = insulin resistance; MBS = metabolic syndrome; NCEP
ATPIII = National Cholesterol Education Program - Adult Treatment
Panel III; PCA = principal components analysis; SBP = systolic blood
pressure; trig. = triglycerides; UCFM = unrestricted confirmatory
factor model; WHO = World Health Organization
\end{sloppypar}
\end{abstract}

\maketitle

\section{Introduction}\label{IntMBS}
Certain risk factors for type 2 diabetes mellitus (DM) and
atherosclerotic cardiovascular disease (CVD) have long been observed
to cluster together in the individual \cite{Mara22, Kyl23}. This
clustering has rendered renewed interest with Reaven's contention
\cite{Rea88, Reav93} that insulin resistance forms its basis. Today,
the complex of interrelated risk factors of metabolic origin is
known as the `metabolic syndrome' (MBS) \cite{EGZ05,Unwi06}. It is
considered to be a major threat to current and future public health,
especially as MBS might result from maladaptive human metabolism in
the face of food energy abundance in combination with a sedentary
lifestyle \cite{WilVo, MCG05}.

Recently, the MBS concept has been hotly debated \cite{Grun05,
KBFS05, Reav05b, Reav05, BE06, LK06, Nara06, AZ08, Gale08, Tenen11}.
A number of interrelated reasons are at the heart of the debate. The
aetiology of the syndrome is largely unknown as to date it is
unclear if a single pathogenetic process promotes the syndrome, or
if multiple different pathogenetic processes need to concur in order
for the syndrome to express itself. Notwithstanding the unknown
aetiology, a number of expert groups, among which the World Health
Organization (WHO) \cite{WHO} and the National Cholesterol Education
Program - Adult Treatment Panel III (NCEP ATPIII) \cite{NCEP}, have
published (slightly) different definitions of MBS intended for
clinical diagnosis. However, clinical evidence on whether MBS is a
better predictor of CVD and DM risk than its individual components,
is equivocal \cite{KBFS05, SNM05}.

The stance on the MBS concept taken here is the following. The
aetiology does not have to be known for the existence of a condition
to be accepted \cite{Gale08}, as is the case with, for example, type
2 DM. Current knowledge is however limited, such that MBS is not
considered a clinical entity, but rather as defining a state of
heightened risk for DM and CVD. The syndromic approach taken is
epidemiological rather than clinical. In the sense that MBS is
deemed to provide a conceptual framework for the clustering of
metabolic risk factors. An important step in furthering
epidemiologic understanding of the syndrome is then an assessment of
the pathophysiological constellation of what are deemed to be
phenotypic expressions of MBS. A constellation we believe to be
driven, not by insulin resistance \cite{Rea88,Reav93}, but by
multiple interrelated metabolic impairments \cite{LMH98}.

Factor analysis (FA) has become an oft-used tool for evaluations of
phenotypic domains underlying MBS \cite{Meigs00}. FA is a
multivariate technique that may reveal a pattern of reduced
dimensionality among a larger set of intercorrelated variables
\cite{Mul10}. FA, however, is often poorly understood while being
routinely executed \cite{Law04}, leading to diffuse findings and
possibly confusing efforts in understanding MBS. This paper aims to
review factor analytic efforts in the study of MBS, with emphasis on
misusage of FA. An alternative factor analytic strategy is proposed
that confronts weaknesses in the application of FA. A high-profile
MBS data set with anthropometric measurements on overweight and
obese children and adolescents is reanalyzed using the alternative
strategy. The findings may give renewed cachet to both FA and its
connection to MBS research.
%
%
\section{Assessing Factor Analytic Efforts in MBS Research}\label{Assess}
\subsection{The Factor Analytic Model}\label{ModFacs}
FA has come to be heavily utilized in the MBS research community
since its seminal usage by Edwards, et al. \cite{EAN94}. An
important question thrusting the FA efforts in MBS research is if a
unifying physiology dominated by insulin resistance underlies the
clustering of metabolic risk variables, or if there are multiple
underlying physiologic phenotypes.

The common factor analytic model assumes that a random
$p$-dimensional vector of observed variables can be grouped by their
covariances or correlations into a lower-dimensional linear
combination of latent variables:
\begin{equation}\label{facM}
\begin{array}{cccccccccc}
  \boldsymbol{\mathrm{z}}_{i}& = &\boldsymbol{\mu} &+&\mathbf{\Lambda} &\cdot& \boldsymbol{\xi}_{i} &+& \boldsymbol{\epsilon}_{i} \\
  (p\times 1) & & (p\times 1) & & (p\times m) & &(m\times 1) & &(p\times 1)
  \end{array}.
\end{equation}
In (\ref{facM}) $\vet{\mathrm{z}}_{i}$ denotes the (possibly
standardized) observed variable of dimension $p$ for person $i$,
$\vet{\mu}$ denotes the intercept, $\vet{\epsilon}_{i}$ denotes the
error measurements for person $i$, and $\mathbf{\Lambda}$ is a
$(p\times m)$-dimensional matrix of factor loadings in which each
element $\lambda_{jk}$ is the loading of the $j$th variable on the
$k$th factor, $j=1,\ldots,p$, $k=1,\ldots,m$. Then $ \vet{\xi}_{i}$
represents a latent variable of dimension $m$, with $m<p$, whose
elements are referred to as common factors. In effect, FA represents
a method of identifying or specifying latent factors that account
for the (co)variances among observed variables by partitioning
observed variance into common variance (attributable to the
underlying latent common factors) and unique variance (among which
are error components) \cite{Mul10}. In the standard model the random
variables $\vet{\mathrm{z}}_{i}$, $\vet{\xi}_{i}$, and
$\vet{\epsilon}_{i}$ are assumed to have Gaussian distributions, although
the development of robust estimation procedures (e.g., \cite{PRC03,LP98})
have somewhat softened the necessity of this assumption.
(See Appendix `Basics of Factor Analysis' for a more detailed overview of the FA
model). The model can have both explorative and confirmative
thrusts. Before assessing these thrusts some general remarks on FA
are made.

A first general comment on the utilization of FA by the MBS research
community is that basic model assumptions are seldom assessed. The
model boasts several implicit assumptions such as a nonsingular
sample covariance matrix and a reasonable proportion of variance
among the observed variables being common variance. The
appropriateness of these assumptions is easily \cite{Kai70} but
rarely assessed. Moreover, the explicit distributional assumptions
imply usage of observed variables of continuous metric and
disqualifies the common FA model for binary and categorical observed
data. Many MBS studies, however, employ standard FA on variables of
non-continuous metric. In such situations extensions of the standard
FA model are needed \cite{Jog07, Mous}.

A second general comment concerns interpretational overextension.
The FA model cannot determine existence of MBS nor assess clinical
importance of MBS as a concept \cite{Law04}. What FA \emph{can} do
is, through the latent factors (when adhering to a realist ontology \cite{Borsboom03}),
give indications of pathophysiological
domains that underlie phenotypic expressions of MBS.

\subsection{Comments on Exploratory Efforts}\label{EXEFF}
In exploratory FA (EFA) both $m$ and the meaning of latent factors
are unknown. In the exploratory sense, FA is a theory-generating
technique used for the identification of meaningful latent factors.
Most MBS studies utilize FA in the exploratory sense
\cite{HKF02,HIB02,Ford03,LPO04,OhHong04,Ang05,Ghossie,CRM06,MMR07,HKT08,KBJ11}.
Many deployments of EFA are however suboptimal.

EFA is often confused with principal components analysis (PCA). PCA
is a data-reductional technique, seeking to identify components. It
resembles the model in (\ref{facM}) without inclusion of error
measurements, leading the components to differ conceptually and
mathematically from latent factors in EFA \cite{Wida}. Components
are weighted linear combinations of observed variables seeking to
efficiently explain observed variance in the data \cite{Jolig},
leaving the explanation of observed covariance secondary
\cite{Wida}. Many MBS studies employing FA, however, seek to obtain
an explanation of the observables' \emph{co}variation through a
small number of explanatory factors. Also, a phenotype is just an
expression of genotype or pathophysiology, indicating the necessity
of including measurement error. As such, employing common EFA would
be more appropriate.

An important decision in EFA is determining the number of factors to
retain. Many methods are heuristical, relying on subjective
judgments or arbitrary cut-off values. The Guttman-Kaiser rule
\cite{Gut54,Kai60} is the most popular rule of thumb. It states that
one should retain at most those factors associated with eigenvalues
whose magnitude exceeds the average eigenvalue (the average
eigenvalue being 1 when using standardized data). This criterion, as
well as many other heuristical factor retention criteria, are prone
to under- (retaining too few factors) and overfactoring (retaining
too many factors) \cite{YG82,Cliff88}.

Given the above it is disconcerting that factor analytic efforts in
MBS research are usually based on what has been termed `Little
jiffy' \cite{Kai70}: Employment of PCA, retainment of components
based on the Guttmann-Kaiser rule followed by a Varimax rotation
\cite{Kai58}, and the subsequent interpretation of rotated
components as if they were common factors. Such mechanical use of
EFA stunts learning and interpretation \cite{PM03}, and is (in part)
responsible for the widely differing results obtained with EFA in
MBS research. See the online supplement to this paper for an overview
of the disparate results obtained with EFA as well as confirmatory FA (next subsection).

\subsection{Comments on Confirmatory Efforts}\label{CONEFF}
Confirmatory FA (CFA) is a theory-testing technique. An \emph{a
priori} factor structure is assumed, with given $m$, with a
pre-specified loadings matrix in which exclusion constraints
indicate which variables are indicators of which latent factor(s),
and with possibly correlated factors and error variances. The model
or models stated then remain to be tested. CFA studies are gaining
interest in MBS research
\cite{STN02,Plade06,SNS06,CF07,Goodm09,BSV09,Esp11}. Standard CFA
however, can also be misapplied.

The evaluation of model fit in CFA is essentially the evaluation of
a diffuse hypothesis as it is unclear in case of misspecification if
the pattern of loadings or the factor dimensionality is to blame
\cite{HD04}. Moreover, specifying a pattern of factor loadings
through exclusion restrictions implies a loss of information in the
sense that more exclusion restrictions are applied than is usually
necessary for identification of the FA model. Additionally,
exclusion restrictions may amount to errors of omission, may make
the unrealistic assumption that items are factorially pure (in the
population), and may induce bias in estimates of the free parameters
\cite{FL00,PK01}. These issues are intricately connected to the
well-known and widespread situation of exploratively obtained factor
structures not being confirmed by CFA \cite{FL00}.

Some recent CFA studies in MBS research claim to provide evidence
that a single latent factor underlies MBS \cite{Plade06,CF07,Esp11}.
These studies include four observed variables, some of which are
functions of several variables usually employed as separate
phenotypic items. This practice (called parceling) is justified by the claim that
utilization of multiple measures for what is believed to be a trait
will lead to a model with multiple factors, thus clouding efforts to
establish a single latent factor underlying MBS. However, the
inclusion of multiple sets of correlated measures does not
irrevocably lead to a model with more than one latent factor, unless
some measures would identify a doublet factor (see Section `Step 1:
Dimensionality Selection' below). Also, the mentioned CFA efforts
may actually provide evidence for a hierarchical latent factor
\cite{STN02} rather than a single pathophysiological domain
underlying MBS, as the usage of functions of variables implies a
(partial) pre-compression of the data. Parceling changes the nature of the data,
may mask model misspecification and thus may inflate goodness-of-fit (indices)
\cite{Marsh13} (cf. \cite{Little13}). Indeed, ``it is possible to argue
(inappropriately) that a single factor solution provides a good fit to
the data under apparently benign parceling strategies" \cite{Marsh13}.
Moreover, while a two-factor
model can be modeled on four variables, such a model would not be
meaningful given that the observables were constructed to represent
a single factor; implying that the one-factor model is the only
model to be meaningfully fitted on the four observed variables. The
possibility to assess if there are multiple (related)
pathophysiological domains underlying phenotypic expressions of MBS
is then denied. A meaningful scientific method, however, allows for
multiple competing theories to be tested \cite{Plat}, as single
hypotheses suffer from confirmation bias.
%
%
\section{An Alternative Factor Analytic Strategy}\label{Alternative}
An alternative confirmatory factor analytic strategy is proposed
that aims to confront the weaknesses in the application of FA. In a
sense the strategy seeks to bridge the EFA -- CFA divide so as to
increase the inferential power of a factor analysis. This strategy
is embedded within the Bayesian model selection approach, whose
analytical advantages have been well documented \cite{Duns01}. The
main Bayesian model selection criterion is the Bayes factor. This
quantity incorporates model fit as well as model complexity and
expresses ``the evidence provided by the data in favor of one
scientific theory, represented by a statistical model, as opposed to
another'' \cite{KR95}. It can be used to compare any two models.
Using Bayes factors, one can compute posterior model probabilities,
which, assuming a uniform prior on the model space, are normalized
Bayes factors. These quantities express model (un)certainty, in the
sense that posterior model probabilities can be interpreted as the
relative amounts of support in the data for the models under
consideration. See Appendix `A Primer on Bayesian Statistics' for
more information on the Bayes factor and
posterior model probabilities.

The strategy proposed consists of the following steps:
\begin{description}
  \item[Step 1]
Determine formally the (intrinsic) number of common factors based on
a weighing of model fit and model complexity. For example along the
lines specified in Peeters \cite{CFWP}, who uses a Bayesian EFA
model for selecting the optimal dimension $m$;
  \item[Step 2]
Whence settled on latent factor dimensionality $m$, specify an
unrestricted confirmatory factor model (UCFM). An UCFM is a FA model
that corresponds to EFA in the sense that only minimal restrictions
are placed on the factor loadings matrix and the factor covariance
matrix for achieving global rotational uniqueness of the factor
solution. However, the restrictions are to be chosen such that they
convey preconceived theoretical meaning and thus render unnecessary
post-hoc rotation of the solution for interpretation purposes.
Peeters \cite{CFWP_Rot} gives minimal conditions for specifying a
UCFM;
  \item[Step 3]
Formulate, using the UCFM obtained in Step 2 as a base model,
competing inequality constrained factor structures making use of
inequality constraints on and between the free parameters in the
loadings matrix. Substantive theory is then not represented by
exclusion restrictions to express a pre-specified factor loading
pattern, but by the imposition of inequality constraints;
\item[Step 4]
Compute the posterior model probability for each constrained model
under consideration and determine the constrained model most
supported by the data.
\end{description}

The strategy explicitly encourages the formulation of competing
inequality constrained theories for (statistical) scrutiny. When
used in full, the thrust of the sequence is confirmatory, with the
explicit separation of dimensionality and pattern selection in order
to avoid embarking on diffuse hypotheses.
%
%
\section{Reanalysis of Data by Weiss et al. \cite{Weiss04}}\label{Pre}
\subsection{Data}\label{DAT}
The data have been described elsewhere \cite{Weiss04}. It considers
a multiethnic, multiracial cohort of 464 nondiabetic obese and
overweight children and adolescents. The data contain measurements
on the body mass index (BMI), blood glucose level at (fasting)
baseline (GB) and two hours after (G2) oral glucose intake (both in
mg/dl), fasting levels of triglycerides (trig.; mg/dl) and
high-density lipoprotein (HDL) cholesterol (mg/dl), systolic and
diastolic blood pressure (SBP, DBP; both in mm Hg), and insulin
resistance (IR). IR was measured through homeostatic model
assessment (HOMA). For more information on the data, see Weiss et
al. \cite{Weiss04}.

Consideration of pediatric samples is important as current
prevalence of MBS among youngsters may give indications of the
future burden of DM and CVD. The measurements are in line with the
American Academy of Clinical Endocrinologists (AACE) position on MBS
\cite{AACE}, which emphasizes the epidemiologic pathophysiological
perspective. The inclusion of IR makes it possible to test theories
regarding the importance of IR in the MBS construct.

The little jiffy approach was the original factor analytic strategy
for analyzing the data \cite{Weiss04}. Here, the alternative
strategy will be utilized for reanalysis. As in Weiss et al.
\cite{Weiss04}, the natural logarithm was taken of the glucose,
insulin resistance and triglycerides measurements to abide the
normality assumption. The data were standardized such that a case of
modeling the correlation matrix is considered. The sample
correlation matrix is nonsingular and the Kaiser-Meyer-Olkin test
\cite{Kai70} indicates that a reasonable proportion of variance
among the variables might be common variance.

\subsection{Step 1: Dimensionality Selection}\label{Step1}
Posterior model probabilities are computed for each model allowed by
the condition $(p-m)^{2}-p-m\geqslant0$. This inequality simply
states that the number of nonredundant elements in the sample
covariance matrix must be greater than or equal to the number of
freely estimable parameters in the model, which places an upper
bound on $m$. The data have eight measured variables $(p=8)$, giving
that the maximum number of factors that can be extracted equals
$m=4$. The computation strategy couples the candidate estimator
method for computing Bayes factors \cite{Chi95} with the use of
training samples \cite{BP96}. This practice allows one to obtain
determinate Bayes factors and subsequent posterior model
probabilities using standard diffuse conjugate or noninformative
priors \cite{Lee07}. The computation strategy is embedded within a
search strategy too weed out models suffering from rank deficiency
in $\mathbf{\Lambda}$ as this is a direct indicator for
overfactoring (see (\ref{identmbs}) in Appendix `Basics of Factor Analysis').
This search excludes the $m=4$ model. The following posterior model
probabilities are obtained when assuming that each model is equally
likely \emph{a priori}: $P(m=1|\mathbf{Z})=0$;
$P(m=2|\mathbf{Z})=1$; and $P(m=3|\mathbf{Z})=0$. The data thus
support the two-factor model.

These results differ from Weiss et al. \cite{Weiss04} and several
other factor analytic efforts in which a three-factor (or higher)
solution was found. A first reason for the retention of more latent
factors in these studies is the tendency of heuristic
factor-selection rules to overfactor \cite{YG82,Cliff88}. More
formal selection procedures, such as the likelihood ratio test in
maximum likelihood EFA and the assessment of information criteria,
do not escape this tendency \cite{LW04,HBY07}.

Another reason for the higher factor solution in other studies might
be the existence of doublet factors. Doublet factors are factors
that arise as the result of common variance due to correlation
between just two variables \cite{Mul10}. Doublet factors are
considered to be conceptually weak factors and the assignment of an
independent latent construct to a doublet factor is contentious. SBP
and DBP are variables that, within the battery of measurements on
phenotypic expressions of MBS, usually correlate only with each
other resulting in a doublet factor to be extracted (usually termed
`hypertension'). The strategy employed here sees past the doublet
factor and indicates the more parsimonious model with two factors as
optimal.

\subsection{Step 2: Base Model Formulation}\label{Step2}
A UCFM for $m=2$ will be formulated for confirmatory efforts.
Abiding conditions given by Peeters \cite{CFWP_Rot}, the following
minimal restrictions on $\mathbf{\Lambda}$ are chosen for global
rotational uniqueness:
\begin{equation}\nonumber
  \mathbf{\Lambda}_{0}=\left[
  \begin{array}{r@{\lambda}l r@{\lambda}l}
    &_{11}             &  &_{12} \\
    &_{21}             &  &_{22} \\
    &_{31}=0           &  &_{32}>0 \\
    &_{41}             &  &_{42} \\
    &_{51}>0           &  &_{52}=0 \\
    &_{61}             &  &_{62} \\
    &_{71}             &  &_{72} \\
    &_{81}             &  &_{82} \\
  \end{array}
  \right]
  \begin{array}{l}
    \mbox{BMI}\\
    \log_{e}\{\mbox{trig.}\}\\
    \mbox{HDL chol.}\\
    \log_{e}\{\mbox{IR}\}\\
    \log_{e}\{\mbox{GB}\}\\
    \log_{e}\{\mbox{G2}\}\\
    \mbox{SBP}\\
    \mbox{DBP}\\
  \end{array}.
\end{equation}
The exclusion restrictions $\{\lambda_{31},\lambda_{52}\}=0$
identify the model up to polarity reversals in the columns. The
polarity truncations $\{\lambda_{32},\lambda_{51}\}>0$ then ensure
global rotational uniqueness of the model. These constraints are
chosen for the following reasons. First, from prior knowledge and
previous analyzes a two-factor solution is deemed to consist of a
glucose and a lipid factor. HDL chol. is then believed to have a
large loading on a lipid (second) factor while having a small
loading on the first factor, and $\log_{e}\{\mbox{GB}\}$ is believed
to have a large loading on a glucose (first) factor while having a
small loading on the second factor. These variables then serve as an
indicator of the respective factors. It is thus reasonable to
specify $\{\lambda_{31},\lambda_{52}\}=0$ and
$\{\lambda_{32},\lambda_{51}\}>0$. Second, the chosen minimal
restrictions comply with all competing inequality constrained
formulations of factor structure to be assessed.

\subsection{Step 3: Formulating Competing Constrained Factor Structures}\label{Step3}
Factor structure for confirmatory efforts is not represented using
exclusion restrictions but by imposing inequality constraints on and
between the parameters left free in the UCFM. The following
inequality constrained competing factor structures are formulated:
\begin{align*}
\mathbf{\Lambda}_{1}&=\left[
  \begin{array}{r@{\lambda}l c r@{\lambda}l}
    &_{11}             & > &  |&_{12}| \\
   |&_{21}|            & < &  -&_{22}  \\
    &_{31}=0           &   &  &_{32}>0 \\
    &_{41}             & > &  |&_{42}| \\
    &_{51}>0           &   &  &_{52}=0 \\
    &_{61}             & > &  |&_{62}| \\
    &_{71}             & > &  |&_{72}| \\
   |&_{81}|            & < &  -&_{82} \\
  \end{array}
  \right]
  \begin{array}{l}
    \mbox{BMI}\\
    \log_{e}\{\mbox{trig.}\}\\
    \mbox{HDL chol.}\\
    \log_{e}\{\mbox{IR}\}\\
    \log_{e}\{\mbox{GB}\}\\
    \log_{e}\{\mbox{G2}\}\\
    \mbox{SBP}\\
    \mbox{DBP}\\
  \end{array},\\ \\
 \mathbf{\Lambda}_{2}&=\left[
  \begin{array}{r@{\lambda}l c r@{\lambda}l}
    |&_{11}|           & < &  -&_{12} \\
    |&_{21}|           & < &  -&_{22}  \\
    &_{31}=0           &   &  &_{32}>0 \\
    &_{41} > .4        &   &  &_{42} < -.4 \\
    &_{51}>0           &   &  &_{52}=0 \\
    &_{61}             & > &  |&_{62}| \\
    &_{71}             & < &  -&_{72} \\
    &_{81}             & < &  -&_{82} \\
  \end{array}
  \right]
   \begin{array}{l}
    \mbox{BMI}\\
    \log_{e}\{\mbox{trig.}\}\\
    \mbox{HDL chol.}\\
    \log_{e}\{\mbox{IR}\}\\
    \log_{e}\{\mbox{GB}\}\\
    \log_{e}\{\mbox{G2}\}\\
    \mbox{SBP}\\
    \mbox{DBP}\\
  \end{array},\\ \\
\mathbf{\Lambda}_{3}&=\left[
  \begin{array}{r@{\lambda}l c r@{\lambda}l}
    &_{11}             & > &  |&_{12}| \\
   |&_{21}|            & < &  -&_{22}  \\
    &_{31}=0           &   &  &_{32}>0 \\
    &_{41}             & > &  |&_{42}| \\
    &_{51}>0           &   &  &_{52}=0 \\
    &_{61}             & > &  |&_{62}| \\
    |&_{71}|< .3       &   &  |&_{72}|< .3   \\
    |&_{81}|< .3       &   &  |&_{82}|< .3 \\
  \end{array}
  \right]
  \begin{array}{l}
    \mbox{BMI}\\
    \log_{e}\{\mbox{trig.}\}\\
    \mbox{HDL chol.}\\
    \log_{e}\{\mbox{IR}\}\\
    \log_{e}\{\mbox{GB}\}\\
    \log_{e}\{\mbox{G2}\}\\
    \mbox{SBP}\\
    \mbox{DBP}\\
  \end{array}.
\end{align*}

A formulation like $\lambda_{71} < -\lambda_{72}$ states that the
negative of $\lambda_{72}$ is believed to be larger than
$\lambda_{71}$. Note that this is a much more informative
formulation than the more usual strategy of setting $\lambda_{71}=0$
and letting $\lambda_{72}$ be free to be estimated in order to
express the belief that SBP is an indicator for the second latent
factor rather than the first one. In the same respect, a formulation
like
\begin{equation}\label{absInq}\nonumber
     \lambda_{61}>|\lambda_{62}|\Rightarrow \left\{\begin{array}{l}
                 \lambda_{61} - \lambda_{62} > 0\\
                 \lambda_{61} + \lambda_{62} > 0
                 \end{array} \right.,
\end{equation}
indicates the belief that $\lambda_{61}$ is larger than
$\lambda_{62}$, irrespective of the latter's sign. A statement like
\begin{equation}\label{absInq2}\nonumber
     |\lambda_{82}|<.3\Rightarrow -.3<\lambda_{82}<.3,
\end{equation}
indicates the belief that $\lambda_{82}$ takes a value in the
interval $[-.3,.3]$.

In all models insulin resistance, blood glucose level at baseline
and two hours after glucose intake are mainly related to the glucose
factor, while fasting levels of triglycerides and HDL cholesterol
form the base of the lipid factor. Note that for the lipid factor
polarity fixation is brought about by demanding $\lambda_{32}>0$.
HDL cholesterol is generally regarded as `good' cholesterol, meaning
that the choice $\lambda_{32}>0$ amounts to modeling a factor
denoting unimpaired lipid metabolism. Hence formulations like
$|\lambda_{21}| < -\lambda_{22}$, as under given polarity truncation
the triglycerides item is believed to be strongly negatively related
to a lipid factor.

Model 1 adds detail to the base model and the basic factors by
stating that BMI and systolic blood pressure are linked to the
glucose factor ($\lambda_{11} > |\lambda_{12}|$, $\lambda_{71} >
|\lambda_{72}|$), while diastolic blood pressure is believed to be
linked to the lipid factor ($|\lambda_{81}| <  -\lambda_{82}$).
Model 2 states the hypothesis that BMI is an indicator for the lipid
rather than the glucose factor ($|\lambda_{11}| < -\lambda_{12}$).
Also, in this model both systolic and diastolic blood pressure are
related to the lipid rather than the glucose factor, with the
additional belief that both blood pressure measures will load
positively on the latter ($\lambda_{71} <  -\lambda_{72}$,
$\lambda_{81} < -\lambda_{82}$). Moreover, the second model states
that insulin resistance may be the measure tying MBS together. In a
multifactor model this would imply that the insulin resistance
measure achieves a large or dominant loading on both the glucose and
lipid factor ($\lambda_{41}> .4$, $\lambda_{42} < -.4$). Model 3
resembles the first model, but states that the association of
systolic and diastolic blood pressure with the factors is rather
loose.

\subsection{Step 4: Constrained-Model Selection and Interpretation}\label{Step4}
Bayes factors for models under inequality constraints are easily
computed \cite{KH07}. Again, standard (diffuse) conjugate and
noninformative priors are utilized. Assuming a uniform prior on the
model space the following posterior model probabilities are obtained
for the constrained two-factor models under consideration:
$P(M_{1}|\mathbf{Z})=.0004$; $P(M_{2}|\mathbf{Z})=0$; and
$P(M_{3}|\mathbf{Z})=.9996$. Conditioned on the data, the third
model receives almost all support.

The third constrained model connects trig. and HDL chol. to a lipid
metabolism factor and IR, GB, and G2 to a glucose metabolism factor.
Moreover, the model states that BMI is related to the glucose
metabolism factor rather than the lipid metabolism factor. The
finding that blood pressure is not an independent pathophysiological
factor is consistent with epidemiologic evidence that insulin
resistance and lipid metabolism play a role in the pathogenesis of
hypertension rather than hypertension being a physiologic phenotype
\cite{LMH98}.

\begin{table}[h]
\caption{Posterior Means and 95\% Credible Intervals for
$\mathbf{\Lambda}_{0}$} \centering
\begin{footnotesize}
\begin{tabular} {ccccccccl}
  \hline\hline\\
    & Factor 1 &   & &  & Factor 2 &  & &\\
    \cline{1-3}\cline{5-7}
  Parameter & Mean & 95\% CI & &Parameter & Mean & 95\% CI & & Item \tabularnewline
    \cline{1-7}\cline{9-9}
  \\
  $\lambda_{11}$ & ~.324 & [ .207,~.440] && $\lambda_{12}$ & -.068 & [-.191, .055]& & BMI \\
  $\lambda_{21}$ & -.006 & [-.303,~.215] && $\lambda_{22}$ & -.653 & [-.956,-.379]& & $\log_{e}$\{trig.\} \\
  $\lambda_{31}$ & -     & -             && $\lambda_{32}$ & ~.706 & [ .442, .940]& & HDL chol. \\
  $\lambda_{41}$ & ~.767 & [ .613,~.921] && $\lambda_{42}$ & -.179 & [-.343,-.022]& & $\log_{e}$\{IR\} \\
  $\lambda_{51}$ & ~.470 & [ .360,~.585] && $\lambda_{52}$ & -     & -            & & $\log_{e}$\{GB\} \\
  $\lambda_{61}$ & ~.355 & [ .205,~.492] && $\lambda_{62}$ & -.124 & [-.289, .036]& & $\log_{e}$\{G2\} \\
  $\lambda_{71}$ & ~.274 & [ .136,~.416] && $\lambda_{72}$ & ~.029 & [-.118, .171]& & SBP \\
  $\lambda_{81}$ & ~.202 & [ .069,~.347] && $\lambda_{82}$ & ~.139 & [-.017, .292]& & DBP \\
  \\
  \hline
\end{tabular}\label{EstMBS}
\end{footnotesize}
\end{table}

The estimates of the UCFM given in Table \ref{EstMBS} also lend
support to model 3. (Confer Appendix `Results Factor Analytic Approach
Weiss et al. \cite{Weiss04}' to see, in contrast,
the parameter estimates obtained with the little jiffy approach of the
original study \cite{Weiss04}). The table contains posterior means and credible
intervals. A credible interval is a posterior probability interval.
For example, the posterior probability that $\lambda_{11}$ lies in
the interval $[ .207,~.440]$ is .95. (See Appendix `Reproduced Correlation Structure'
for an indication of the success of the two-factor UCFM in
retrieving the sample correlation matrix).

The estimates indicate that the hypertension variables seem to be
relatively weak in their association with the respective factors.
The credibility intervals indicate that these variables are mostly
related to the glucose metabolism factor, which would be in line
with the hypothesis that hypertension is related to insulin
resistance and impaired glucose metabolism \cite{Rea88,Reav93}.
Regarding the contention that insulin resistance is the basis for
MBS: IR is related to both the glucose and lipid factors. However,
the posterior mean of the loading tying IR to the lipid metabolism
factor is relatively small and the upper bound of its credibility
interval approaches zero. The second inequality constrained model is
thus rightly not supported by the data.

The two latent factors are appreciably correlated with a posterior
mean of $-.277$ and a 95\% credible interval of $[-.417,-.137]$.
Note that, as stated in Section `Step 3: Formulating Competing Constrained
Factor Structures', the second factor
models unimpaired lipid metabolism. Thus, the selected model
indicates that, given the data and measurements, impaired glucose
metabolism and impaired lipid metabolism are positively related
pathophysiological domains. Moreover, the two factors connect to two
main hypotheses regarding syndrome aetiology, stating that the
risk-factor associations are due to abnormality of the
insulin/glucose metabolism and/or abnormality of the lipid
metabolism \cite{LMH98}.
%
%
\section{Discussion}\label{Pre}
Applications of FA in MBS research were evaluated. Lacunae in both
EFA and CFA were discussed. It is argued that the mechanical use of
EFA and misunderstandings of CFA are, at least in part, responsible
for the widely differing results obtained with FA on data with
phenotypic expressions of MBS.

An alternative factor analytic strategy is proposed. The strategy
consists of four steps and aims to confront the weaknesses in
application of the FA model, by: (i) Formally assessing optimal
choice of factor dimensionality; (ii) Canceling the need for
post-hoc rotation of the factor solution; (iii) Allowing to express
a confirmatory factor structure through  informative inequality
constraints rather than through rigid exclusion restrictions; (iv)
Encouraging the formulation of competing inequality constrained
theory-based expressions of factor structure, in order to avoid
confirmation bias.

The alternative strategy was utilized in reanalyzing a high-profile
data set on which factor analyzes were previously employed. The data
consider eight variables as phenotypic expressions of MBS in a
cohort of nondiabetic overweight and obese children and adolescents
\cite{Weiss04}. The reanalysis based on the alternative strategy
implied a more parsimonious constellation of pathophysiological
domains underlying phenotypic expressions of MBS than the original
analysis (and many other analyzes). The selected two-factor solution
stresses correlated factors of impaired glucose metabolism and
impaired lipid metabolism. This solution does not assign
hypertension a separate factor which is consistent with
epidemiologic evidence that insulin resistance and lipid metabolism
play a role in the pathogenesis of hypertension rather than
hypertension being a physiologic phenotype \cite{LMH98}. Moreover,
there is no strong evidence of insulin resistance being dominant in
both the glucose and lipid domains.

Several limitations of this study should be considered. First, the
data consider a multiethnic cohort while MBS may express itself
differently across ethnic groups and gender. Note, however, that the
specification of a factor model through inequality constraints would
also be helpful in assessing measurement (factorial) invariance
across groups. Second, MBS may develop with age and with the advent
of DM and CVD, implying that the data might represent a snap-shot of
phenotypic expressions related to MBS. Third, the proposed factor
analytic strategy is more involved than routine uses of EFA and
regular CFA in the sense that it requires more computation time and
puts higher cognitive demands on the researcher. Nevertheless, these
drawbacks are felt to be outweighed by the advantages of the
strategy and the proposed steps are deemed to form a viable analytic
alternative for other studies seeking to use FA.

The findings suggest that there are two correlated pathophysiological
domains underlying the phenotypic expressions of MBS included in the
analysis. These domains are characterized by impaired glucose
metabolism and impaired lipid metabolism, respectively. These
findings indirectly point to the possibility that several different
pathogenic processes need to coincide in order to be able to
identify a MBS construct. It might be timely to postulate the
possible existence of a multifactorial aetiology.

\section*{Acknowledgements}
This work was
supported by grant NWO-VICI-453-05-002 of the Netherlands
Organization for Scientific Research (NWO). It was written while the author was a
Ph.D. candidate at the Department of Methodology and Statistics, Utrecht University,
Utrecht, the Netherlands and is part of the authors' Ph.D. thesis. This version
is a postprint of: Peeters, C.F.W., Dziura, J., \& van Wesel, F. (2014). 
Pathophysiological Domains Underlying the Metabolic Syndrome: An Alternative
Factor Analytic Strategy. Annals of Epidemiology, 24: 762--770.
This postprint is released under a Creative Commons BY-NC-ND license. 

\appendix
\section*{Appendix: Basics of Factor Analysis}\label{FacBasic}
The unrestricted factor model is considered. Let
$\mathbf{Z}^{\mathrm{T}}\equiv[\vet{\mathrm{z}}_{1},\ldots,\vet{\mathrm{z}}_{n}]$
define (standardized) $p\mbox{-variate}$ observation vectors on
$i=1,\ldots,n$ subjects, such that
$\vet{\mathrm{z}}_{i}^{\mathrm{T}}\equiv[z_{i1},\ldots,z_{ip}]\in\mathbb{R}^{p}$
denotes a realization of the random vector
$Z_{i}^{\mathrm{T}}\equiv[Z_{i1},\ldots,Z_{ip}]\in\mathbb{R}^{p}$.
Also, let
$\mathbf{\Xi}^{\mathrm{T}}\equiv[\vet{\xi}_{1},\ldots,\vet{\xi}_{n}]$
define $m$-variate vectors of latent factor scores on $n$ subjects
with
$\vet{\xi}_{i}^{\mathrm{T}}\equiv[\xi_{i1},\ldots,\xi_{im}]\in\mathbb{R}^{m}$.

The model (\ref{facM}) maintains the following assumptions: (i)
$\vet{\mathrm{z}}_{i}\ci \vet{\mathrm{z}}_{i'},\forall i\neq i'$;
(ii) rank$(\mathbf{\Lambda})=m$; (iii) $\vet{\epsilon}_{i}\sim
\mathcal{N}_{p}(\vet{0}, \vet{\Psi})$, with $\vet{\Psi}\equiv
\mbox{diag}(\psi_{11},\ldots, \psi_{pp})$, and $\psi_{jj}>0$; (iv)
$\vet{\xi}_{i}\sim \mathcal{N}_{m}(\vet{0}, \mathbf{\Phi})$; and (v)
$\vet{\xi}_{i}\ci\vet{\epsilon}_{i'},\forall i,i'$. The likelihood
for the observations conditional on the realization of
$\mathbf{\Xi}$ can then be expressed as:
\begin{align}\label{likelihood}\nonumber
  L(\vet{\mu},\mathbf{\Lambda}, \mathbf{\Xi}, \mathbf{\Psi}, \mathbf{\Phi}; \mathbf{Z}) &=
  \prod_{i=1}^{n}f(\vet{\mathrm{z}}_{i}|\vet{\mu},\mathbf{\Lambda},\vet{\xi}_{i},\mathbf{\Psi}, \mathbf{\Phi})\\
  & =
  \prod_{i=1}^{n}(2\pi)^{-\frac{p}{2}}|\mathbf{\Psi}|^{-\frac{1}{2}}
  \exp\left\{-\frac{1}{2}\vet{\epsilon}_{i}^{\mathrm{T}}\mathbf{\Psi}^{-1}\vet{\epsilon}_{i}\right\},
\end{align}
where
$\vet{\epsilon}_{i}=\vet{\mathrm{z}}_{i}-\vet{\mu}-\mathbf{\Lambda}\vet{\xi}_{i}$.
Marginalizing over $\vet{\xi}_{i}$ the likelihood of the observed
data can be obtained:
\begin{align}\label{likelihoodMarg}\nonumber
  L(\vet{\mu},\mathbf{\Lambda}, \mathbf{\Psi}, \mathbf{\Phi}; \mathbf{Z})&=
  \prod_{i=1}^{n}\int f(\vet{\mathrm{z}}_{i}|\vet{\mu},\mathbf{\Lambda},\vet{\xi}_{i},\mathbf{\Psi},\mathbf{\Phi})
  g(\vet{\xi}_{i}|\mathbf{\Phi})\,\partial\vet{\xi}_{i}\\
  &=\prod_{i=1}^{n}(2\pi)^{-\frac{p}{2}}|\mathbf{\Lambda}\mathbf{\Phi}\mathbf{\Lambda}^{\mathrm{T}}+\mathbf{\Psi}|^{-\frac{1}{2}}\\\nonumber
  &~~~~~~~~\times\exp\left\{-\frac{1}{2}(\vet{\mathrm{z}}_{i}-\vet{\mu})^{\mathrm{T}}[\mathbf{\Lambda}\mathbf{\Phi}\mathbf{\Lambda}^{\mathrm{T}}+\mathbf{\Psi}]^{-1}
  (\vet{\mathrm{z}}_{i}-\vet{\mu})\right\},
\end{align}
giving that the factor decomposition constrains the covariance
structure of the $\vet{\mathrm{z}}_{i}$ to
\begin{equation}\label{Fundamental}
\mathbf{\Sigma}=\mathbf{\Lambda}\mathbf{\Phi}\mathbf{\Lambda}^{\mathrm{T}}+\mathbf{\Psi}.
\end{equation}
Then, for existence (vi), generally $(p-m)^{2}-p-m\geqslant0$,
simply stating that the number of nonredundant elements in the
sample correlation matrix $\mathbf{S}$ must be greater than or equal
to the number of freely estimable parameters in $\vet{\Sigma}$,
which places an upper bound on $m$.

Now, $\mathbf{\Phi}\in\mathbb{R}^{m\times m}$ denotes the factor
covariance matrix, giving that (\ref{Fundamental}) represents an
oblique model in which the latents may share covariation. Note that,
for positive definite $\mathbf{\Phi}$, we may always find
$\mathbf{V}\in\mathbb{R}^{m\times m}$ such that $\mathbf{\Phi} =
\mathbf{V}\mathbf{V}^{\mathrm{T}}$, and
\begin{equation}\label{ident3mbs}
  \mathbf{\Sigma}=\mathbf{\Lambda}\mathbf{\Phi}\mathbf{\Lambda}^{\mathrm{T}}+\mathbf{\Psi} =
  (\vet{\Lambda}\mathbf{V})[\mathbf{V}^{-1}\mathbf{\Phi}(\mathbf{V}^{-1})^{\mathrm{T}}](
  \vet{\Lambda}\mathbf{V})^{\mathrm{T}}+\vet{\Psi}=
  (\vet{\Lambda}\mathbf{V})(\vet{\Lambda}\mathbf{V})^{\mathrm{T}}+\vet{\Psi}.
\end{equation}
Equation (\ref{ident3mbs}) implies that any oblique representation
has equivalent orthogonal representations. The orthogonal
representation makes the following statements on identification less
involved.

It is well known that for given $\mathbf{\Lambda}$ and
$\mathbf{\Psi}$, the former is defined uniquely only up to rotation.
Correspondingly the FA literature has focussed mainly on
identification of $\mathbf{\Psi}$. The main result of which is that
if assumption (vi) holds, $\mathbf{\Psi}$ is almost surely
identified \cite{BMW94}. This result is however contingent upon the
rank of $\mathbf{\Lambda}$. The implications of a failure to abide
model assumption (ii) were explored by Anderson and Rubin
\cite{AR56} and Geweke and Singleton \cite{GS80}. Suppose that
rank$(\mathbf{\Lambda})=r<m$. Then there exists a matrix
$\mathbf{Q}\in\mathbb{R}^{m\times (m-r)}$ for which
$(\mathbf{\Lambda}\mathbf{V})\mathbf{Q}=\boldsymbol{0}$ and
$\mathbf{Q}^{\mathrm{T}}\mathbf{Q}=\mathbf{I}_{m-r}$, such that for
any $\mathbf{M}\in\mathbb{R}^{p\times (m-r)}$ with mutually
orthogonal rows
\begin{equation}\label{identmbs}
  \mathbf{\Sigma}=(\vet{\Lambda}\mathbf{V})(\vet{\Lambda}\mathbf{V})^{\mathrm{T}}+\vet{\Psi} =
  (\mathbf{\Lambda}\mathbf{V}+\mathbf{M}\mathbf{Q}^{\mathrm{T}})
  (\mathbf{\Lambda}\mathbf{V}+\mathbf{M}\mathbf{Q}^{\mathrm{T}})^{\mathrm{T}}+
  (\mathbf{\Psi}-\mathbf{M}\mathbf{M}^{\mathrm{T}}).
\end{equation}
Equation (\ref{identmbs}) implies that no consistent estimator of
$\mathbf{\Psi}$ exists if $\mathbf{\Lambda}$ fails to be of full
column rank. This may induce corresponding multimodalities in the
densities of $\mathbf{\Psi}$ and $\mathbf{\Lambda}$ \cite{LW04}, and
is related to the choice of factor dimensionality and the
possibility of retaining too many factors.

The FA model also copes with an inherent indeterminacy of the
parameters, being: Rotational indeterminacy of the factor solution.
Assume that $\mathbf{R}\in\mathbb{R}^{m\times m}$ is an arbitrary
nonsingular matrix. Returning to the implied covariance structure of
the observed data, we then have
\begin{equation}\label{ident2mbs}
  \mathbf{\Sigma}=\mathbf{\Lambda}\mathbf{\Phi}\mathbf{\Lambda}^{\mathrm{T}}+\mathbf{\Psi}
  =(\mathbf{\Lambda}\mathbf{R})[\mathbf{R}^{-1}\mathbf{\Phi}(\mathbf{R}^{\mathrm{T}})^{-1}]
(\mathbf{\Lambda}\mathbf{R})^{\mathrm{T}}+\mathbf{\Psi},
\end{equation}
implying that there is an infinite number of alternative matrices
$\mathbf{\Lambda}^{\ddag}=\mathbf{\Lambda}\mathbf{R}$ and
$\mathbf{\Phi}^{\ddag}=\mathbf{R}^{-1}\mathbf{\Phi}(\mathbf{R}^{\mathrm{T}})^{-1}$
that generate the same covariance structure $\mathbf{\Sigma}$. The
operation $\mathbf{\Lambda}\mapsto\mathbf{\Lambda}\mathbf{R}$ is
termed `rotation'. Thus, in any solution, $\mathbf{\Lambda}$ can be
made to satisfy $m^{2}$ additional conditions, which is naturally
equivalent to the number of independent elements of $\mathbf{R}$.

From the above it is clear that any method of estimation requires at
a minimum $m^{2}$ restrictions on $\mathbf{\Lambda}$ and
$\mathbf{\Phi}$. The EFA tradition usually achieves this by
requiring that $\mathbf{\Phi}=\mathbf{I}_{m}$ and
$\mathbf{\Lambda}^{\mathrm{T}}\mathbf{\Psi}^{-1}\mathbf{\Lambda}$ be
diagonal accompanied by an order condition on the diagonal elements.
These restrictions are arbitrary such that whence estimation is
settled EFA traditionally endeavors on applying a rotation that
satisfies certain criteria for interpretation purposes. Peeters
\cite{CFWP_Rot} has given minimal conditions for the formulation of
a rotationally unique confirmatory unrestricted factor model that
cancels the need for post-hoc rotation.

\section*{Appendix: A Primer on Bayesian Statistics}\label{BayesPrimer}
The Bayesian viewpoint is distinct from the classical approach to
statistics. Let $\mathbf{\Theta}$ denote a model-specific collection
of unknown parameters of continuous metric. The frequentist approach
solely utilizes the likelihood of the observed data
$L(\mathbf{\Theta};\mathbf{X})$, in that a retrospective evaluation
is made of a certain statistic used to estimate $\mathbf{\Theta}$
over all possible $\mathbf{X}$ values conditional on the true
unknown $\mathbf{\Theta}$ which is deemed fixed. The Bayesian
approach views $\mathbf{\Theta}$ as random. This allows for
probability statements about $\pi(\mathbf{\Theta}|\mathbf{X})$, the
distribution of model parameters conditioned on the observed data.
To provide the mentioned conditional probabilities, a joint
probability function for $\mathbf{\Theta}$ and $\mathbf{X}$ must be
provided for. To this purpose a prior distribution
$\pi(\mathbf{\Theta})$ must be specified, which reflects the
formalized knowledge or uncertainty about the parameters before
observation of the data. Using a basic property of conditional
probability known as Bayes' rule \cite{Bayes,Laplace}, one obtains
the posterior distribution as:
\begin{equation}\label{bayes}
\pi(\mathbf{\Theta}|\mathbf{X})=\frac{L(\mathbf{\Theta};\mathbf{X})\pi(\mathbf{\Theta})}
{\displaystyle\int
L(\mathbf{\Theta};\mathbf{X})\pi(\mathbf{\Theta})\,\partial\mathbf{\Theta}}.
\end{equation}
Expression (\ref{bayes}) encapsulates the core machinery of Bayesian
statistics, whose flexibility has proven to extend to complex
problems (consult, for example \cite{Press03,GCSR}). The denominator
in (\ref{bayes}) is called the prior predictive density or marginal
likelihood and is key in Bayesian model selection.

Let us shortly review Bayesian model selection for latent variable
models. Let $\boldsymbol{\vartheta}$ denote latent data. For the
factor model described in Section `The Factor Analytic Model',
$\mathbf{\Theta}=\{\boldsymbol{\mu},\mathbf{\Lambda},\mathbf{\Psi},\mathbf{\Phi}\}$
and $\boldsymbol{\vartheta}=\mathbf{\Xi}$. Now, let
$g(\boldsymbol{\vartheta}|\mathbf{\Theta})$ denote the density of
latent data $\boldsymbol{\vartheta}$ given $\mathbf{\Theta}$ and
assume that the complete data likelihood consists of
$L(\mathbf{\Theta},\boldsymbol{\vartheta};
\mathbf{X})g(\boldsymbol{\vartheta}|\mathbf{\Theta})$. Suppose also
that the prior $\pi(\mathbf{\Theta})$ is available for the unknown
model parameters $\mathbf{\Theta}$. The marginal likelihood is then
expressed as:
\begin{equation}\label{margNonintro}
  m(\mathbf{X})=\int L(\mathbf{\Theta},\boldsymbol{\vartheta};\mathbf{X})
  \pi(\mathbf{\Theta})g(\boldsymbol{\vartheta}|\mathbf{\Theta})
  ~\partial(\mathbf{\Theta},\boldsymbol{\vartheta}).
\end{equation}
The marginal likelihood expresses the likelihood of the data
conditional on the model entertained. This quantity is of import in
the construction of the Bayes factor, the main Bayesian model
selection criterion. Suppose that $S$ competing models $M_{s}$ are
under consideration, for $s=1,\ldots, S$. The Bayes factor of
$M_{s}$ to $M_{s'}$ is then expressed as \cite{Jef35, Jef61}:
\begin{equation}\label{BF1intro}
  B_{ss'}=\frac{m_{s}(\mathbf{X})}{m_{s'}(\mathbf{X})}=\frac{\displaystyle\int
  L_{s}(\mathbf{\Theta}_{s}, \boldsymbol{\vartheta}_{s};\mathbf{X})
  \pi_{s}(\mathbf{\Theta}_{s})g_{s}(\boldsymbol{\vartheta}_{s}|\mathbf{\Theta}_{s})
  ~\partial(\mathbf{\Theta}_{s},\boldsymbol{\vartheta}_{s})}
  {\displaystyle\int
  L_{s'}(\mathbf{\Theta}_{s'}, \boldsymbol{\vartheta}_{s'};\mathbf{X})
  \pi_{s'}(\mathbf{\Theta}_{s'})g_{s'}(\boldsymbol{\vartheta}_{s'}|\mathbf{\Theta}_{s'})
  ~\partial(\mathbf{\Theta}_{s'},\boldsymbol{\vartheta}_{s'})}.
\end{equation}
The Bayes factor embodies the ratio of posterior odds to prior odds
for the models under consideration. The expression in
(\ref{BF1intro}) resembles a likelihood ratio. But instead of
evaluating the respective likelihoods at the maximum likelihood
estimates, the parameters are integrated out with respect to the priors.
The BF thus can be viewed as representing a
`weighted' likelihood ratio that provides a measure ``of the
evidence provided by the data in favor of one scientific theory,
represented by a statistical model, as opposed to another''
\cite{KR95}. The Bayes factor behaves like a natural Occam's razor,
as model fit and complexity are accounted for in the marginal
likelihood \cite{SS82,JB92}. For interpretation the quantity
(\ref{BF1intro}) can be referred to half-units on the $\log_{10}$
scale (Appendix B of \cite{Jef61}) or one can consider
$2\log_{e}B_{ss'}$ \cite{KR95}, which is on the same scale as
likelihood ratio statistics. Another interpretational aid might be
the posterior model probability, defined as \cite{BP96}:
\begin{equation}\label{PMP}
P(M_{s'}|\mathbf{X})=\left(\sum_{s=1}^{S}\frac{p_{s}}{p_{s'}}\cdot
B_{ss'}\right)^{-1}.
\end{equation}
In (\ref{PMP}) $p_{s}$ denotes the prior probability one assigns to
model $M_{s}$ being best, $\sum_{s=1}^{S}p_{s}=1$. The posterior
model probability $P(M_{s'}|\mathbf{X})$ gives the posterior
probability, given the batch of models under consideration, that
model $M_{s'}$ is the correct model for the data at hand. A
normalization of the Bayes factor ensues when letting $p_{s}=S^{-1}
~\forall s$.

The Bayes factor as a model selection criterion has several
advantages (cf., \cite{KR95,Lee07}): (i) It provides both a measure
of evidence against a competing model and a measure of support for
the alternative model; (ii) It will not by default favor the
alternative model in (very) large samples; (iii) It allows one to take model uncertainty into account,
thus providing a consistent quantity for the comparison of a
multitude of competing models; (iv) It can handle the comparison of
both nested and nonnested models.
For an introduction to Bayesian statistics see \cite{Press03,GCSR}. For
Bayesian factor analysis see \cite{Lee07, MuthenBFA}. For
inequality-constrained-model selection for Bayesian FA as well as
Bayesian treatments of both EFA and CFA, see \cite{CFWP}.

\section*{Appendix: Results Factor Analytic Approach Weiss et al. \cite{Weiss04}}\label{WeissResults}
Table \ref{LJdata} contains the results of the little jiffy approach
to the data as given in the original study (Table 3 in \cite{Weiss04}).
A three-component solution was obtained in which the first component was interpreted as `obesity and glucose metabolism',
the second as `dyslipidemia', and the third as `blood pressure'.

\begin{table}[h]
\caption{Results little jiffy analysis on the data
\cite{Weiss04}} \centering
\begin{tabular} {ccccl}
  \hline\hline\\
    & Components &   & & \\
    \cline{1-3}
  1 & 2 & 3 & & Item \tabularnewline
    \cline{1-3}\cline{5-5}
  \\
  ~.44 & ~.13 & ~.06 & & BMI \\
  ~.09 & ~.83 & ~.04 & & $\log_{e}$\{trig.\} \\
  -.13 & -.82 & ~.06 & & HDL chol. \\
  ~.76 & ~.27 & ~.15 & & $\log_{e}$\{IR\} \\
  ~.72 & -.14 & ~.07 & & $\log_{e}$\{GB\} \\
  ~.67 & ~.10 & -.06 & & $\log_{e}$\{G2\} \\
  ~.15 & ~.09 & ~.79 & & SBP \\
  ~.01 & ~.09 & ~.83 & & DBP \\
  \\
  \hline
\end{tabular}\label{LJdata}
\end{table}

\section*{Appendix: Reproduced Correlation Structure}\label{CorStruc}
Table \ref{SCM} contains the sample correlations, the correlations
reproduced by the factor analysis, and the residual correlations
(sample correlation minus reproduced correlation).

\begin{table}[h!]
\caption{Matrix containing observed (Pearson), reproduced, and
residual correlations} \centering
\begin{tabular} {lrrrrrrrr}
  \hline\hline
  \tabularnewline
                       & 1 & 2 & 3 & 4 & 5 & 6 & 7 & 8 \tabularnewline
  \hline
  \\
  1 BMI                  &                &        &               &                &        &              &        &   \\
  ~~~~~observed           & 1.000         &        &               &                &        &              &        &   \\
  ~~~~~reproduced         & 1.002         &        &               &                &        &              &        &   \\
  ~~~~~residual           & -.002         &        &               &                &        &              &        &   \\
  2 $\log_{e}$\{trig.\}  &                &        &               &                &        &              &        &   \\
  ~~~~~observed           & .041           & 1.000  &               &                &        &              &        &   \\
  ~~~~~reproduced         & .101           & .984   &               &                &        &              &        &   \\
  ~~~~~residual           & -.060          & .016   &               &                &        &              &        &   \\
  3 HDL chol.            &                &        &               &                &        &              &        &   \\
  ~~~~~observed           & -.143          & -.423  & 1.000         &                &        &              &        &   \\
  ~~~~~reproduced         & -.111          & -.459  & .998          &                &        &              &        &   \\
  ~~~~~residual           & -.032          & .036   & .002          &                &        &              &        &   \\
  4 $\log_{e}$\{IR\}     &                &        &               &                &        &              &        &   \\
  ~~~~~observed           & .314           & .259   & -.254         & 1.000          &        &              &        &   \\
  ~~~~~reproduced         & .291           & .250   & -.276         & 1.006          &        &              &        &   \\
  ~~~~~residual           & .023           & .009   & .022          & -.006          &        &              &        &   \\
  5 $\log_{e}$\{GB\}     &                &        &               &                &        &              &        &   \\
  ~~~~~observed           & .071           & .051   & -.066         & .387           & 1.000 &              &        &   \\
  ~~~~~reproduced         & .161           & .082   & -.092         & .384           & 1.004 &              &        &   \\
  ~~~~~residual           & -.090          & -.031  & .026          & .003           & -.004 &              &        &   \\
  6 $\log_{e}$\{G2\}     &                &        &               &                &        &              &        &   \\
  ~~~~~observed           & .121           & .197   & -.105         & .336           & .233   & 1.000        &        &   \\
  ~~~~~reproduced         & .141           & .143   & -.157         & .338           & .183   & .999         &        &   \\
  ~~~~~residual           & -.020          & .054   &  .052         & -.002          & .050   & .001         &        &   \\
  7 SBP                  &                &        &               &                &        &              &        &   \\
  ~~~~~observed           & .132           & .065   & -.028         & .190           & .104   & .084         & 1.000  &   \\
  ~~~~~reproduced         & .089           & .029   & -.033         & .212           & .125   & .100         & .999   &   \\
  ~~~~~residual           & .043           & .036   & .005          & -.022          & -.021  & -.016        & .001   &   \\
  8 DBP                  &                &        &               &                &        &              &        &   \\
  ~~~~~observed           & -.013          & -.016  & .084          & .093           & .095   & -.010        & .332   & 1.000 \\
  ~~~~~reproduced         &  .047          & -.056  & .059          & .111           & .077   &  .048        & .047   & .999 \\
  ~~~~~residual           & -.060          & .004   & .025          & -.018          & .018   & -.058        & .275   & .001 \\
  \\
  \hline
\end{tabular}\label{SCM}
\end{table}

\bibliographystyle{vancouver}
\bibliography{ReferenceMBS_Revision}

\addresseshere

\cleardoublepage

\renewcommand{\theequation}{S\arabic{equation}}
\renewcommand{\thefigure}{S\arabic{figure}}
\renewcommand{\thetable}{S\arabic{table}}
\renewcommand{\thesection}{\arabic{section}}

\setcounter{section}{0}
\setcounter{subsection}{0}
\setcounter{equation}{0}
\setcounter{figure}{0}
\setcounter{table}{0}
\setcounter{page}{1}

\phantomsection
\addcontentsline{toc}{section}{Supplementary Material}
\begin{center}
{\huge SUPPLEMENTARY MATERIAL}
\end{center}

\bigskip
\bigskip
Tables \ref{OverviewCA} and \ref{OverviewAdditional} below list overviews of published studies on the MBS employing factor analytic techniques. Table \ref{OverviewCA} lists studies among the child and adolescent cohort. Table \ref{OverviewAdditional} lists studies among subjects other than children and adolescents. Studies were found using the PubMed search engine \cite{PubMed} by pairing the search term ``factor analysis" with each in \{``metabolic syndrome", ``insulin resistance syndrome", and ``syndrome X"\} (the names that are regularly used to refer to the MBS). By usage of these terms English-language publications were sought with (online) publication dates from 1994 (year of publication of the seminal article by \cite{EAN94}) to January 2014. See \cite{PMP06, FL08} and the online supplement to \cite{Plade06} for additional (complementary) overview tables. Please consider the remarks below for full understanding of Tables \ref{OverviewCA} and \ref{OverviewAdditional}.

\bigskip\noindent
\textbf{Eligibility criteria.} Included studies focus their factor analytic efforts, as in the main text, on variables of a metabolic nature. Studies that focus mainly on hemostatic, inflammatory, lifestyle, or diet variables were excluded. Studies that provided no information on their factor analytic procedure (extraction, retention, and rotation) were also excluded. In addition, studies that considered binary or categorical variables in standard factor models were also not considered.

\bigskip\noindent
\textbf{Data extraction.} If a study that focuses on the mentioned variables for exclusion in addition ran a factor analytic procedure on the traditional metabolic variables, then only the results of the analysis on the metabolic variables are reported. When study and validation cohorts were available in a certain study, only the study cohort is reported when results comply. Also, when FA was performed on various (sub)sets of variables, we chose to report the FA on the combination of items most in concurrence with the data (re-)analyzed in this study.

When representing EFA efforts (including PCA) the description of the factors is based on the loading cut-offs chosen by the authors of the original studies. When representing CFA efforts in the tables below, we focus on the factor structure, not on other model details such as correlated error variables. When a higher-order factor model is utilized in a certain study only the first-order factors are described (the second-order factor is always termed `MBS'). It is however explicitly indicated when a higher-order model is fitted. In addition, we only report final models when describing CFA efforts. Details such as usage of modification indices can be found in the studies themselves.

The assessment of the stability of the factor analytic model is generally known as the issue of `measurement invariance'. This issue arises in longitudinal data (is the model stable over the measured time-points) or when the total sample is considered to consist of subgroups (is the model stable over all subgroups that make up the total sample). Invariance can be studied on the level of the model structure (number of factors and factor structure) or, given a certain model structure, on the level of the parameter values. A formal approach would be to perform multiple group analyses, possibly paired with a hierarchy of invariance tests (see e.g., \cite{Bollen}). This formal approach is strongly tied to CFA. An informal approach, usually performed in EFA-type analyses, would be to perform separate EFA's (or PCA's) on the data from the respective subgroups or time-points and to assess loosely if the model is stable. In the tables below we use `assessment of measurement invariance' to refer to the more formal approach while we use `subgroup analyzes' to refer to the more informal approach. In the tables below it is explicitly indicated if, within a certain study, factor analytic results differ over the various subgroups or time-points. If the model structure is considered invariant (in the original study) this will be designated with `structure considered invariant'. If model structure and parameter values are considered invariant (in the original study) this will be designated with `model considered invariant'.

\bigskip\noindent
\textbf{Reading the tables.} In the tables below `approach' refers to the factor analytic approach taken and `factors' refers to either common factors (in the case of a true factor analytic approach) or principal components (in case of a PCA approach). The sample size given is the size of the sample included in the factor analytic efforts. Characteristics refer to the characteristics of the sample included in the FA.

Note that in the description of factors the ordering is of import in the PCA approach and in an EFA approach with certain rotation criteria. The factors then can be understood as being in descending order of (percent of) variance explained (in the observed variables). In a CFA approach, the stated ordering of factors is arbitrary. The numbering in this latter case is solely to convey the number of factors modelled.

The reported studies are ordered according to year of publication. While the metabolic variables may at first seem to differ widely between the included studies, one may note that in many cases surrogate measures are used. For example, often the apolipoproteins B and A-I are used as surrogates for the LDL and HDL cholesterol fractions, respectively. Note also that the measures for insulin resistance can vary over studies, e.g., fasting and postload insulin, HOMA-IR, or the intravenous glucose tolerance test are some of the measures that are oft-used. In addition, the tables use `/' to designate a ratio, e.g., trig./HDL chol. indicates the triglycerides over HDL cholesterol ratio.

The tables utilize, next to the abbreviations used in the main text, the following additional abbreviations:

\begin{sloppypar}
\begin{description}
  \item
  \%BF = percent body fat; 17HP = 17-hydroxyprogesterone; Adi = adiponectin; Apo A-I = apolipoprotein A-I; Apo B = apolipoprotein B; AST = abdominal skinfold thickness; BW = body weight; CASPIAN = Childhood and Adolescence Surveillance and Prevention of Adult Non-Communicable Disease; FFFA = fasting free fatty acids; FG = fasting glucose; FI = fasting insulin; fib. = fibrinogen; FT = free testosterone; HbA1c = glycated hemoglobin; HC = hip circumference; HDL$_2$ = high density lipoprotein 2 cholesterol; HDL tot. = total high density lipoprotein cholesterol; IAF = (CT-measured) intra-abdominal fat area; IAI = Instituto Auxologico Italiano; IMGD = insulin-mediated glucose disposal; IS = insulin sensitivity; IVGTT = intravenous glucose tolerance test; LDL chol. = high density lipoprotein cholesterol; LDL-PPD = low-density lipoprotein peak particle diameter; LR = likelihood ratio; MAP = mean arterial pressure; ML = maximum likelihood; NFG = non fasting glucose; NHLBI = National Heart, Lung, and Blood Institute; NO$_x$ = nitric oxide metabolites; PAI-1 = plasminogen activator inhibitor-1; PAL = physical activity level (minutes/week); PCDD/Fs = polychlorinated dibenzo-p-dioxin, dibenzofurans persistant organic pollutants; PFFA = postload free fatty acids; PG = postload glucose; PI = postload insulin; PIn = ponderal index; RA = renin activity; SSK = subscapular skinfold; SSPG = steady state plasma glucose; S:T = subscapular to triceps; SuRFNCD = Surveys of Risk Factors of Non-Communicable Diseases; TC = total cholesterol; TER = trunk extremity ratio; TFM = trunk fat mass; tria. = triacylglycerols; TSI = Torres Strait Islander; TTTS = trunk-to-total skinfolds; UA = uric acid; U:C = urinary albumin - creatinine ratio; WBC = white blood cell count; WC = waist circumference; WHR = waist-to-hip ratio.
\end{description}
\end{sloppypar}

\fontsize{5pt}{6pt}\selectfont
\begin{center}
\begin{landscape}
\begin{longtable}{l|l|l|l}
\caption[Factor analyzes of the MBS in the child and adolescent cohort]{Factor analyzes of the MBS in the child and adolescent cohort} \label{OverviewCA} \\

\hline \multicolumn{1}{l|}{Study} &
\multicolumn{1}{l|}{Characterization} &
\multicolumn{1}{l|}{Approach} &
\multicolumn{1}{|l}{Description factors} \\ \hline
\endfirsthead

\multicolumn{4}{c}%
{{\tablename\ \thetable{} (Continued)}} \\
\hline
\multicolumn{1}{l|}{Study} &
\multicolumn{1}{l|}{Characterization} &
\multicolumn{1}{l|}{Approach} &
\multicolumn{1}{|l}{Description factors} \\ \hline
\endhead

  Chen \emph{et al.} \cite{Chen1999} & - Bogalusa Heart Study                               & Little jiffy  & 1. trig., HDL chol., FG, FI, PIn   \\
                                     & - Biracial (black/white) community-based population  &               & 2. SBP, DBP, FI \\
                                     & - $n = 4,522$ children, adolescents, and young adults&               &  \\
  &  &  &  \\
  &  &  &  \\
  Chen \emph{et al.} \cite{Chen2001} & - Bogalusa Heart Study                               & Little jiffy, ethnic & \emph{Whites} \\
                                     & - Biracial (black/white) community-based population  & subgroup analyzes    & 1. \%BF, SBP, DBP   \\
                                     & - $n = 264$ (132 whites, 132 African Americans)      &                      & 2. \%BF, HDL chol., trig. \\
\hline
\end{longtable}

\begin{longtable}{l|l|l|l}
\caption[Factor analyzes of the MBS in additional
cohorts]{Factor analyzes of the MBS in additional
cohorts} \label{OverviewAdditional} \\

\hline \multicolumn{1}{l|}{Study} &
\multicolumn{1}{l|}{Characterization} &
\multicolumn{1}{l|}{Approach} &
\multicolumn{1}{|l}{Description factors} \\ \hline
\endfirsthead

\multicolumn{4}{c}%
{{\tablename\ \thetable{} (Continued)}} \\
\hline
\multicolumn{1}{l|}{Study} &
\multicolumn{1}{l|}{Characterization} &
\multicolumn{1}{l|}{Approach} &
\multicolumn{1}{|l}{Description factors} \\ \hline
\endhead

  Edwards \emph{et al.} \cite{EAN94,Edwards1997}& - Kaiser-Permanente Women Twins Study & Little jiffy  & 1. BW, WC, FI, FG   \\
                                                & - Unrelated, nondiabetic women        &               & 2. FI, PI, FG, PG, SBP \\
                                                & - $n = 281$                           &               & 3. trig., HDL chol., LDL-PPD               \\
   &  &  &   \\
   &  &  &   \\
  Donahue \emph{et al.} \cite{Donahue1997}  & - Miami Community Health Study                            & Little jiffy  & 1. UA, SBP, DBP, HDL chol., trig., WC, IMGD \\
                                            & - Nondiabetic African-Americans and non-Hispanic whites   &               & 2. FI, FG, DBP, IMGD\\
                                            & - $n = 50$                                                &               &      \\
\hline
\end{longtable}
\end{landscape}
\end{center}

\end{document}